\documentstyle[11pt,aaspp4]{article}
\singlespace

\received{19 October 1998}
\accepted{26 January 1999}
\slugcomment{Astrophysical Journal, in press}

\lefthead{Cotton et al.}
\righthead{Parsec Scale Radio Jet in NGC315}

\begin{document}

\title{A Parsec Scale Accelerating Radio Jet\\
    in the Giant Radio Galaxy NGC315\\}

\author{W. D. Cotton\altaffilmark{1}}
\affil{National Radio Astronomy Observatory, 520 Edgemont Rd.,
Charlottesville VA 22903-2475}

\author{L. Feretti and G. Giovannini\altaffilmark{2}}
\affil{Istituto di Radioastronomia del CNR, Via P. Gobetti 101, I-40129
Bologna, Italy.}

\author{L. Lara}
\affil{Instituto de Astrof\'{\i}sica de Andaluc\'{\i}a, CSIC, Apdo. 3004, 
18080 Granada, Spain.}

\and

\author{T. Venturi}
\affil{Istituto di Radioastronomia del CNR, Via P. Gobetti 101, I-40129
Bologna, Italy.}

% Notice that each of these authors has alternate affiliations, which
% are identified by the \altaffilmark after each name.  The actual alternate
% affiliation information is typeset in footnotes at the bottom of the
% first page, and the text itself is specified in \altaffiltext commands.
% There is a separate \altaffiltext for each alternate affiliation
% indicated above.

\altaffiltext{1}{NRAO is operated by Associated Universities, Inc.\ under cooperative
agreement with the National Science Foundation.} 
\altaffiltext{2}{Dipartimento di Fisica, Universit\'a di Bologna, via B. Pichat 6/2, 
I-40127 Bologna, Italy.}

% The abstract environment prints out the receipt and acceptance dates
% if they are relevant for the journal style.  For the aasms style, they
% will print out as horizontal rules for the editorial staff to type
% on, so long as the author does not include \received and \accepted
% commands.  This should not be done, since \received and \accepted dates
% are not known to the author.

\begin{abstract}
Observations of the core of the giant radio galaxy NGC315 made with VLBI
interferometers are discussed in the context of a 
relativistic jet.  The sidedness asymmetry  suggests Doppler favoritism from a
relativistic jet.  
The presence of moving features in the jet as well as
jet counter--jet brightness ratios hint at an accelerating,
relativistic jet. An increasing jet velocity is also supported by a comparison
of the jet's observed properties with the predictions of an adiabatic expansion
model.
On the parsec scale, the jet is unpolarized at a wavelength of 6 cm to
a very high degree in clear distinction to the high polarization seen
on the kiloparsec scale.
\end{abstract}

% The different journals have different requirements for keywords.  The
% keywords.apj file, found on aas.org in the pubs/aastex-misc directory, 
% contains a list of keywords used with the ApJ and Letters.  These are 
% usually assigned by the editor, but authors may include them in their 
% manuscripts if they wish. 

%\keywords{globular clusters,peanut clusters,bosons,bozos}

% That's it for the front matter.  On to the main body of the paper.
% We'll only put in tutorial remarks at the beginning of each section
% so you can see entire sections together.

% In the first two sections, you should notice the use of the LaTeX \cite
% command to identify citations.  The citations are tied to the
% reference list via symbolic KEYs.  We have chosen the first three
% characters of the first author's name plus the last two numeral of the
% year of publication.  The corresponding reference has a \bibitem
% command in the reference list below.
%
% Please see the AASTeX manual for a more complete discussion on how to make
% \cite-\bibitem work for you.   

\section{Introduction}

There is growing evidence that the jets observed
in the inner regions of powerful extragalactic radio source  have
highly relativistic flows (see e.g. \cite{urry95}).
The relativistic effects of such jets will cause their apparent
properties to be strongly dependent on the viewing angle.
The lines of argument for relativistic motion include
side--to--side asymmetries in brightness 
and polarization, prominence of the core, less X-ray emission than
expected from a very high brightness radio core, and apparent
superluminal motions in the radio jets.
The case for highly relativistic motions in low luminosity sources is
less clear. 
At present, evidence is growing that parsec scale jets are
relativistic also in low power sources (see Lara et al., 1997), but a
detailed study of more sources is necessary to understand the jet
dynamics. 
In recent years, we have been engaged in a program of studying a
complete sample of low power radio sources in order to address this
question; this is the ninth paper of this series.

The radio source associated with the galaxy NGC~315 
has been well studied since its initial discovery (\cite{davis67}), and
is classified as
a low luminosity Fanaroff-Riley type 1 (FR I) radio galaxy.
This galaxy is a 12th magnitude
dusty elliptical (\cite{ebneter85}) at z=0.0167 with 
H$\alpha$ + [NII] emission lines (\cite{marcha96}). 
The large
scale (50') structure of this radio source (0055+30) has been
observed using Arecibo, Westerbork and the VLA
(\cite{fanti76,bridle76,willis81,jagers87}). 
On this large scale, the radio source has a long, straight, highly
polarized jet (\cite{fomalont80,jagers87}) with a sharp bend
near the end.  
The counter-jet is much weaker and has an ``S'' shaped symmetric bend at its end.
The source contains a prominent, peaked spectrum core concident with the
nucleus of the galaxy.
This core has been observed using VLBI arrays by \cite{linfield81,preuss83},
and recently by \cite{venturi93} at 18, 6 and 3.6 cm.
On the basis of high core prominence, low level of nuclear X-ray emission
as well as the jet sidedness, \cite{venturi93} (but see also
\cite{lara97}) concluded  that the jet was highly relativistic, but
could not detect the expected motions in the jet in comparison with
the images  of \cite{linfield81} and \cite{preuss83}.
The flux density of the core was monitored from 1974 to 1980 by
\cite{ekers83}, but no variations were detected.

At the distance of this source, 1 milliarcsecond (mas) corresponds to
0.47 pc.
\footnote {H$_0$ = 50 km sec$^{-1}$Mpc$^{-1}$ is assumed 
throughout the paper.}
We present new observations of NGC~315 using the Very Long Baseline
Array (VLBA) radio telescope.
Many of these data include measurements of the linear polarization.

\section{Observations and Data Reduction}

In this section the details of the observations and the reduction
techniques are given.
The resultant images are presented in the following section.

The galaxy NGC~315 was observed at wavelengths of 6 and 3.6 cm  ( 5
and 8.4 GHz) using the VLBA on 1994 November 14 (total intensity
only), and at 6 cm on 1995  October 28 (dual polarization), 1996 May
10 (dual polarization) and 1996 October 7 (dual polarization). 
These observations are summarized in Table \ref{VLBAobs}.
In the November 1994 observations we switched wavelength every half an hour
between 6 and 3.6 cm,
obtaining a good uv-coverage at both wavelengths.
In the 1995 October observations, the VLA failed to give fringes to the
VLBA although internal VLA observations showed no problems.  
The loss of the VLA seriously reduced the sensitivity which is needed for
linear polarization observations and the observations were repeated in
1996 May.
During these re--observations, the complete failure of VLBA-St. Croix and
serious tape recorder problems at VLBA-Pie Town and VLBA-Mauna Kea
 reduced the resolution of
the data and the observations were rescheduled for 1996 October.
This latter session was successful and produced the polarization
measurements presented here.

   The VLBA data were all correlated on the VLBA correlator in
Socorro (NM, USA) and all processing used the NRAO AIPS package.
Amplitude calibration was initially done using the standard method
employing measured system temperatures and an assumed sensitivity
calibration. 
The VLBA calibration for the VLA was determined from the source to
system temperature ratio derived from the VLA data.
In the 1995 and 1996 observations,
 the amplitude calibration was
refined using the measured flux densities of the compact calibrator
sources (0235+164 in 1995, BL Lac in 1996).

   Phase calibration of the 1995 and 1996 data employed the pulsed
phase calibration system to remove variations in the difference of the
phases of the right-- and left--handed polarized signals.
All data were globally fringe fitted (\cite{schwab83}) and then self
calibrated.  

The VLA was used in parallel with the 1995 and 1996 VLBA observations
to determine the flux density and polarization of the VLBA calibrators.
Because NGC315 is too extended to be used as a phasing calibrator for the
VLA, the nearby source 0042+233 was used to periodically phase the VLA
antennas. 
In the 1996 October session, 3C48 was used to calibrate the flux
density and polarization angle and 3C84 was used to determine the
instrumental polarization for the VLA.

\begin{table} [t]
\begin{center}
\caption{VLBA Observations}
\medskip
\begin{tabular}{lllc} 
\hline \hline 

Date &  Stations &  Polarization & BW (MHz) \\
\hline
1994 November 14 & PT,KP,LA,FD,NL,BR,OV,HN,MK,SC & single & 32 \\
1995 October 28  & PT,KP,LA,FD,NL,BR,OV,HN,MK,SC & dual circular & 64 \\
1996 May 10      & Y,PT,KP,LA,FD,NL,BR,OV,HN,MK & dual circular & 64 \\
1996 October 7  & Y,PT,KP,LA,FD,NL,BR,OV,HN,MK,SC & dual circular & 64 \\
\hline
\label{VLBAobs}
\end{tabular}
\medskip \\

\end{center} 
\tablecomments{
1994 November 14 at 6 and 3.6 cm.; other epochs only at 6 cm.;\\
Y: Phased VLA; PT: VLBA-Pie Town; 
KP: VLBA-Kitt Peak; LA: VLBA-Los Alamos; 
FD: VLBA-Fort Davis; NL: VLBA-North Liberty; 
BR: VLBA-Brewster; OV: VLBA-Owens Valley;
HN: VLBA-Hancock; MK: VLBA-Mauna Kea;
SC: VLBA-St. Croix.
}
\end{table}

The polarization calibration and imaging followed the general method of
Cotton (1993).
The polarization calibrator for the 1995 observations was 0235+164
and for the 1996 observations, BL Lac.
%The VLA and VLBA observations of these sources are summarized in Table
%\ref{polcal}.
The 1996 October data was of much higher quality than the previous
measurements, so the polarization results presented here are from
that session.

Observations of BL Lac were used to derive the cross polarized delay and
phase corrections to the VLBA data.
The polarization angle corrections were derived independently in each
8 MHz band using the sum of the Q and U CLEAN components from a
deconvolution.
This procedure makes the plausible assumption that all of the source
measured by the VLA was also measured by the VLBA.
Instrumental polarization corrections for the VLBA were initially
determined from measurements of 3C84 which was assumed to be
unpolarized. 
After it was determined that NGC315 had no detectable polarization
with this calibration, corrections to the VLBA instrumental
polarization calibration were determined from the NGC315 data assuming
the source to be unpolarized in the region to which these data were
sensitive. 

%\placetable{polcal}

\section{Results}
\subsection{Monitoring of the arcsecond core flux density}
   Flux density monitoring of NGC315 at a wavelength of 6 cm by \cite{ekers83}
from 1974 to 1980 with the WSRT  showed no evidence for variations. 
They found a core flux density of 633 mJy with an rms = 27 mJy. 
More recent VLA observations show a flare between 1990 and 1995.
Flux density measurements at 6 cm are given in Table \ref{vary}.
Many of these measurements are given in this paper for the first time
as reported in Table \ref{vary}.
They result from long integrations on the VLA which have been
extensively cross calibrated; relative errors are likely lower than
the quoted precision.  
More details on the new VLA data will be given in a future paper, in
preparation, on the large scale structure of NGC315. 
The flux density flare is confirmed also by two observations at 3.6
cm. that give a core flux density of 588 mJy in 1990.92 (Venturi et 
al., 1993) and of 746 mJy on June 1994 (present paper).

\begin{table} [t]
\begin{center}
\caption{NGC315 Core Flux Density at 6 cm}
\medskip
\begin{tabular}{lcc} 
\hline \hline 

Date &  Flux Density & Reference \\
     & mJy \\
\hline
1978.48 & 620.0 & 1 \\
%% 1980.82 & 555.0 & 2 \\  attention to reference numbers!
1989.28 & 585.9 & 2 \\
1995.83 & 735.2 & 3 \\
1996.36 & 694.8 & 3 \\
1996.77 & 686.2 & 3 \\
1996.84 & 668.1 & 3 \\
1997.53 & 688.7 & 3 \\
\hline
\label{vary}
\end{tabular}
\medskip \\
\end{center}  
\tablecomments{
The quoted flux densities for the  1989-1997 measurements have a
relative accuracy of better than 0.1 mJy.\hfill\break
{\em References:}
1=Bridle et al. 1979, 
%2=Rudnick et al. 1980, 
2=Venturi et al. 1993, 3=this paper. 
}
\end{table}

\subsection{Total intensity}

   Calibrated and edited visibility data were used to produce total intensity
maps. We used the AIPS package following the standard procedure: a
first map was made using the AIPS task IMAGR and several iteration of
phase self-calibration followed by a final phase and gain self-calibration
were made.
In order to exploit the high dynamic range in the 1996 October data,
baseline dependent complex gain factors were determined from the
NGC315 data by dividing by the Fourier transform of the best self
calibrated model and averaging over the entire time of the observations.
Since the NGC315 jet is very homogeneous,
special care was taken to avoid CLEAN artifacts by using a very
low gain factor (0.03) in all the deconvolutions.

\subsubsection{Total intensity images}
\begin{figure}
\plotone{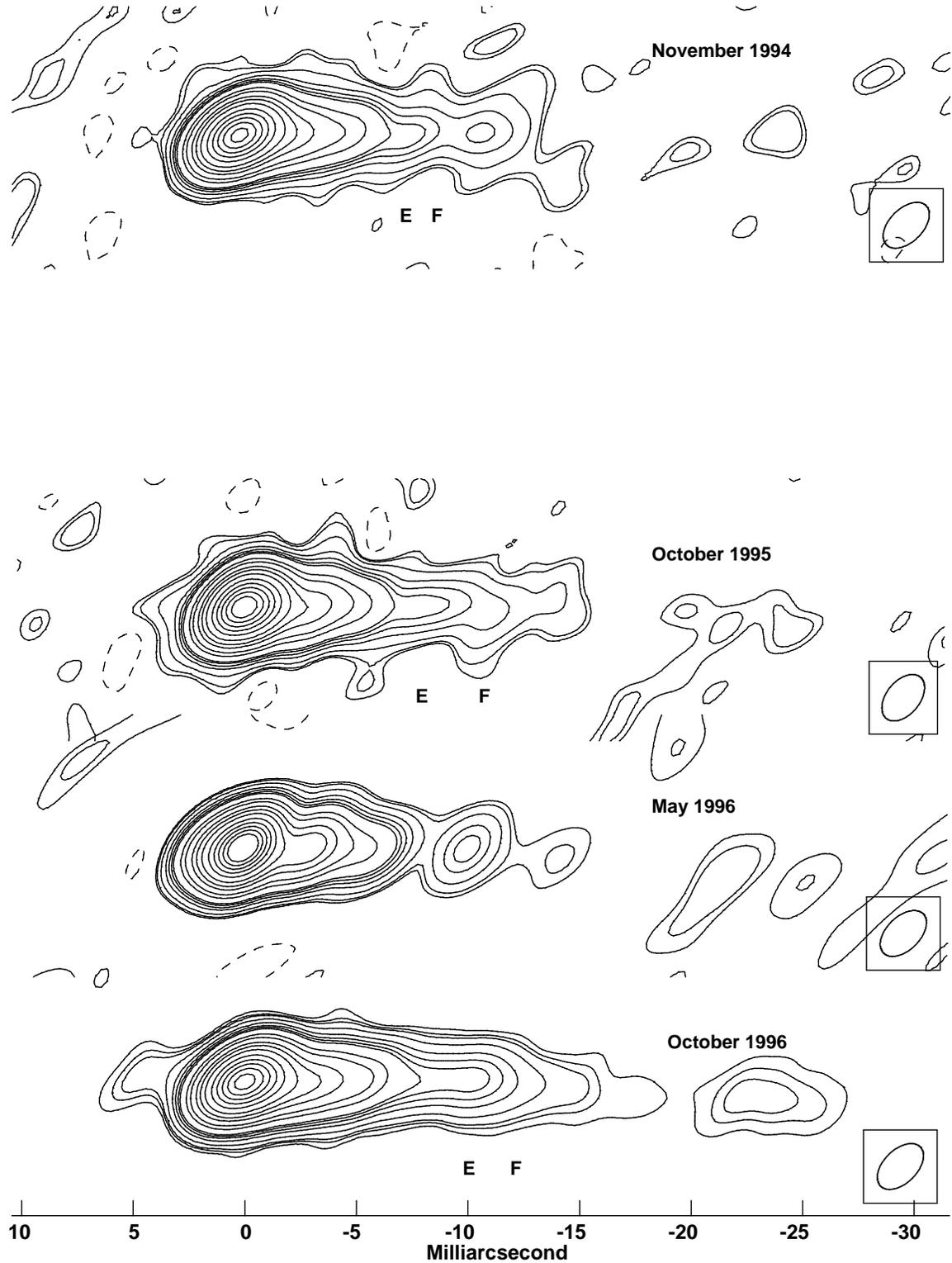}
\caption{
The 6 cm images at 4 epochs shown with the same resolution, on the same
scale and with similar contour levels.
The vertical spacing gives roughly the temporal separation of the
images. 
Labels E and F indicates the location of features.
Contour levels are shown at -1, 0.7, 1, 2, 4, 6, 7, 10, 20, 30, 40,
70, 100, 150, 200, 200, 300 and 350 mJy/beam.
The 1996 October image has an additional contour at 0.35 mJy/beam.
The noise level is 0.4 mJy in the first 2 epochs, 0.5 mJy in the May
1996 map and 0.04 mJy in the last epoch.
The images have been rotated on the sky by -41.5$^\circ$;
the size of the restoring beam (2.5 $\times$ 1.5 mas in position angle
-4$^\circ$) is shown in the lower right corner. 
\label{N315Cont}}
\end{figure}

   In Fig. \ref{N315Cont}, we present the NGC 315 maps at 6 cm at the
four different epochs; details of the images are given in the
figure caption. 
All the images have been rotated by -41.5$^\circ$  and convolved to the
same angular resolution. 
The radio source morphology, a strong core emission and a straight
jet, are in good agreement in all four different images. 
The noise level in the first three images is similar, although
the lack of three telescopes (see Sect. 2) in the May 1996 image
produced some residual artifacts and a more irregular zero level in
this epoch. 
Due to the good quality of the data and to the very accurate
calibration,  the October 1996 image has a noise level a factor 10
lower than the others and a very low level of artifacts.
%Fig. \ref{N315XCcomp} a) shows the total intensity image
%obtained from November 1994 data at 3.6 cm. 

\begin{figure}
\plotone{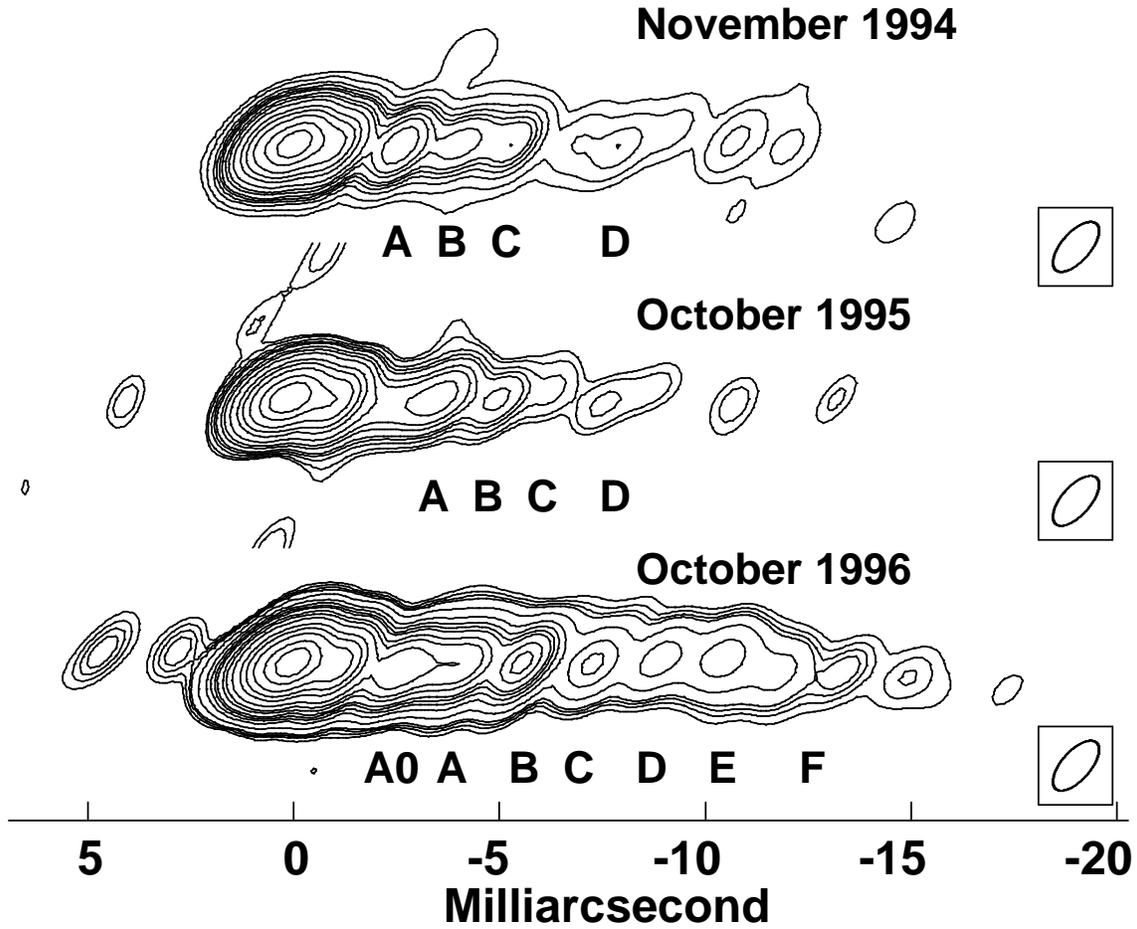}
\caption{
The 6 cm images at 3 epochs shown superresolved to the same
resolution as the 1994 3.6 cm image and shown on the same
scale and with similar contour levels.
Labels A0 -- F indicates the location of features.
Contour levels are shown at 1, 2, 4, 6, 8, 10, 15, 20, 30, 50, 70, 100,
150 and 200 mJy/beam.  The lowest contour in the 1995 October image is
at 3 mJy/beam and the 1996 November image has additional contours at 0.4,
0.6, and 0.8 mJy/beam. 
The images have been rotated on the sky by -41.5$^\circ$;
the size of the restoring beam (1.5 $\times$ 0.7 mas in position angle
0$^\circ$) is shown in the lower right corner. 
\label{N315HRCont}}
\end{figure}

\begin{figure}
\plotone{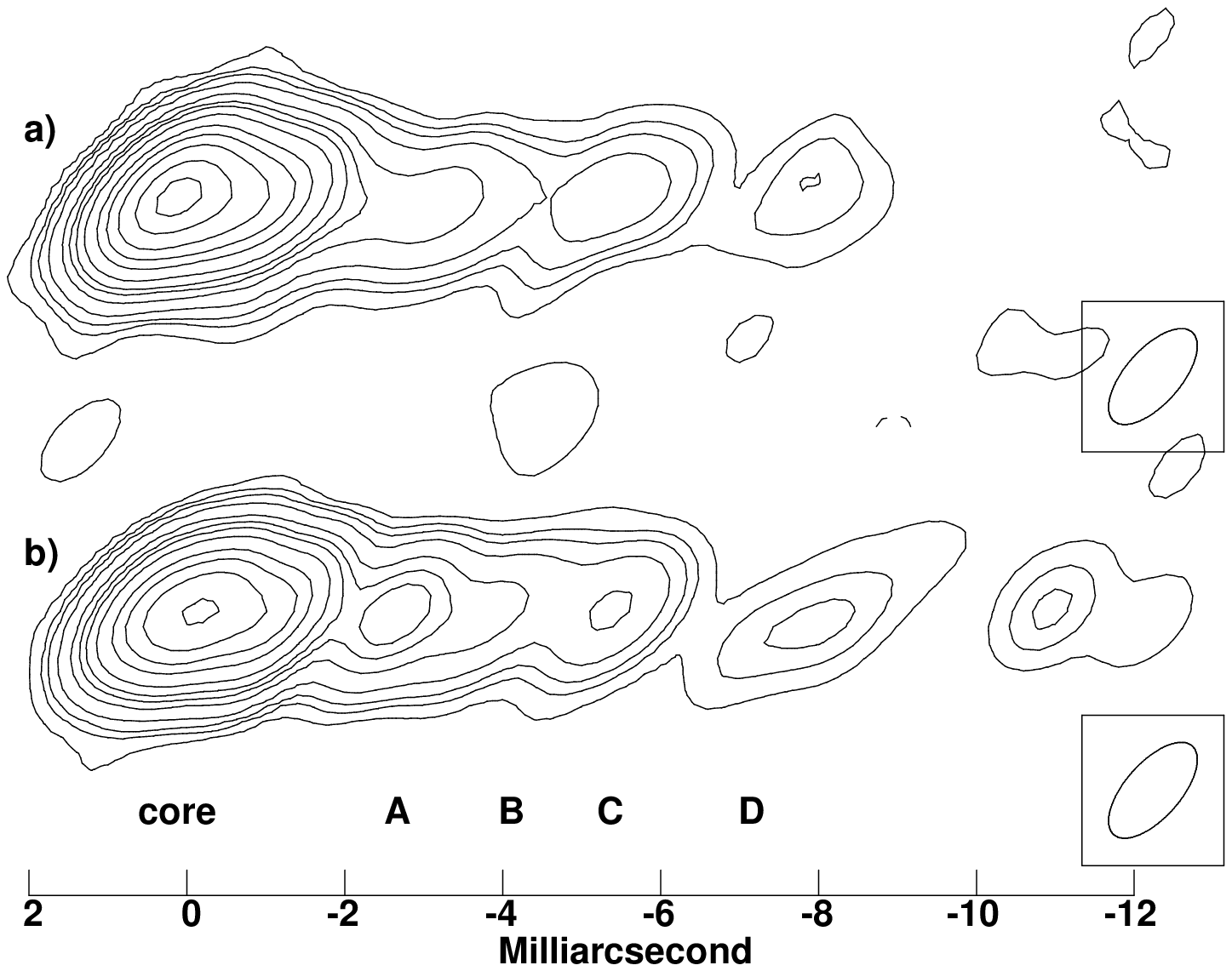}
\caption{
a) The 3.6 cm image of NGC~315 from 1994 November.
The peak is 345 mJy/beam and contour levels are shown at -1.5, 1.5, 3, 4.5,
7.5, 15, 22.5, 30, 45, 75, 105, 150, 225 and 300 mJy/beam.
The image has been rotated on the sky by -41.5$^\circ$;
the size of the restoring beam (1.5 $\times$ 0.7 mas in position angle
0$^\circ$) is shown in the lower right corner. \hfill\break
b) The 6 cm image of NGC~315 from 1994 November super-resolved to the
resolution shown in a).
The peak is 243 mJy and contour levels are the same as in a).
The position scale is from the 3.6 cm image.
The locations of features discussed in the text are indicated.
\label{N315XCcomp}}
\end{figure}

A comparison of the images in Figure \ref{N315Cont} suggests  evidence
for expansion in the jet. 
To investigate further the possibility of motions in the jet, enhanced
resolution images were obtained from the 6 cm images by restoring the
CLEAN components from deconvolution with a Gaussian of the size used
for the 3.6 cm image.
This was not done for the 1996 May observations due to the lack of
long baselines.
These high resolution images are presented in Fig. \ref{N315HRCont}
similar to those in Fig. \ref{N315Cont}. 
To check if the Clean deconvolution could have created artifacts in the super
resolved maps, we compared the November 94 super--resolved map at 6 cm with
the same epoch 3.6 cm map (Fig. \ref{N315XCcomp}a,b).
The strong similarities between the two images of NGC~315 
give some confidence in the validity of our procedure. 

\subsubsection{ Spectral index image}

   We used the same epoch observations (November 1994) at 6 and 3.6 cm
to produce a spectral index image of the NGC315 jet at the resolution
of 2.5 $\times$ 1.5 mas.  
The two images were produced using as similar as possible uv--coverage,
with the same gridding and convolved with the same beam.
%We decided not to align them using the peak flux since the self-absorbed
%core could have a slightly different position at the two frequencies
%and we aligned the image using the well defined peak visible at about 10
%mas from the core. 
The calibration procedure used on this data does not result in
accurate relative registration of the 6 and 3.6 cm images.
As the location of the peak of the ``core'' may be frequency
dependent, a weaker but more isolated feature ``C'' from the images
shown in Fig. \ref{N315XCcomp} was used to align the images. 
With this registration, the peaks of the core at the two wavelengths
are not coincident, but slightly shifted ( 0.3 mas) with the 3.6 cm
peak ``upstream'' of the 6 cm peak.
In the following analysis, the center of activity is assumed to be at
the location of the 3.6 cm peak.

\begin{figure}
\plotone{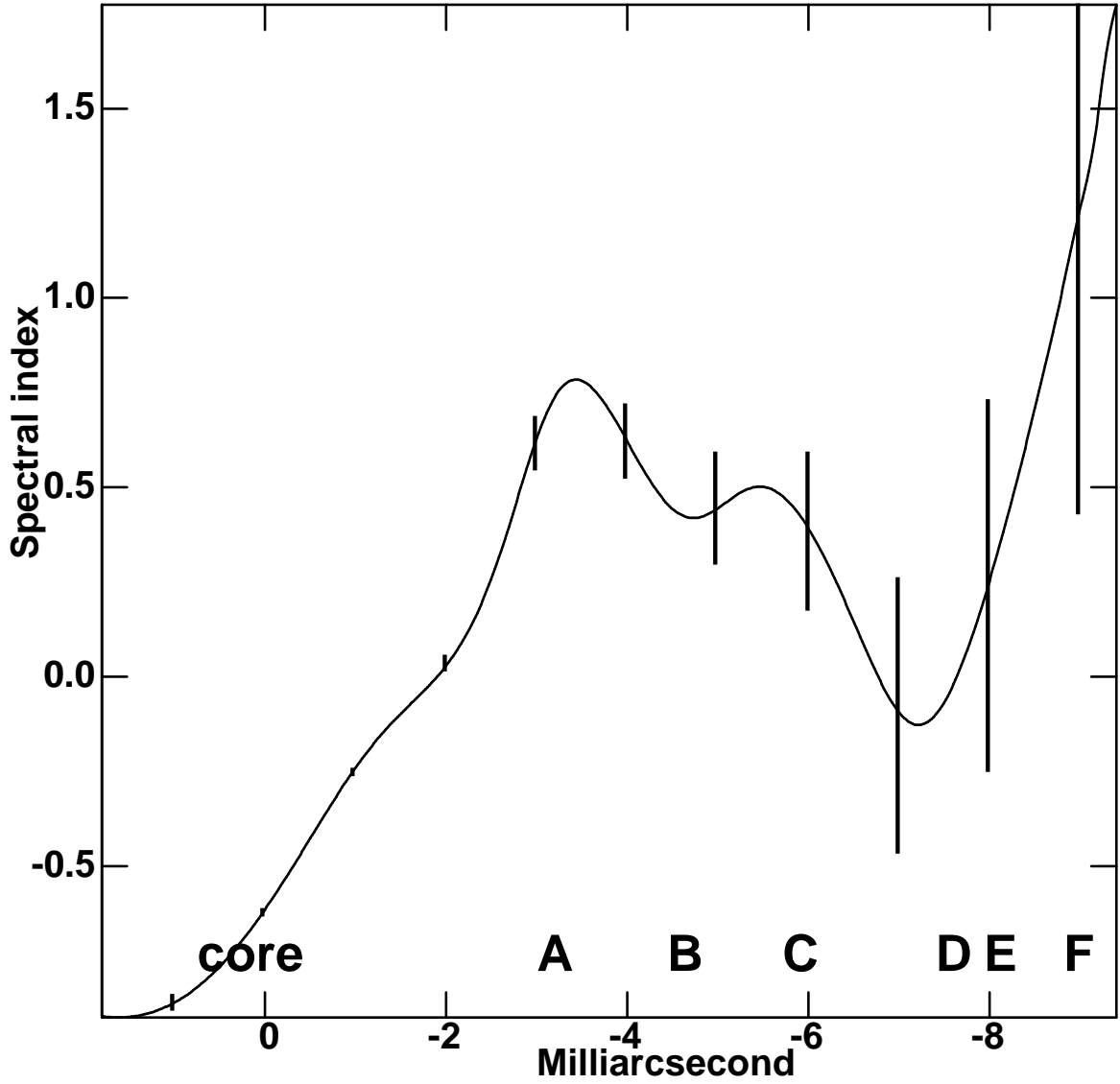}
\caption{
The spectral index ($S = S_0\nu^{-\alpha}$) between 6 and 3.6 cm along the
ridge line of the jet derived from the 1994 November observations.
Resolution used is that of the 6 cm data (2.5 $\times$ 1.5 mas).
The locations of features discussed in the text are marked as well as
the error bars at selected locations.
\label{SpIndex}}
\end{figure}

   The trend of the spectral index along the ridge of maximum brightness
is shown in Fig. \ref{SpIndex}. 
The core region appears strongly self-absorbed and has an inverted
spectrum. 
The spectral index of the jet away from the core is consistent with a
relatively constant value of approximately 0.5.
%The jet region from the core up to 10 mas shows a steepening of the
%spectrum with distance from the core. 
%We have regions with a steep spectrum with $\alpha$ from 0.3 to 2.0
%with three secondary minima which corresponds to marginally evident
%knots in the total intensity image.  
%We note that in anycase the general trend of the spectral index along
%the jet is to increase moving from the core to the exthernal region. 
%The strong minima show that the marginal evident knots in the total
%intensity data are real and that a spectral index image is able to
%show up them better than a total intensity one at the same
%resolution.
%Superimposed on the general steepening of the jet are regions of
%flatter spectrum coinciding with the locations of the apparent
%features in the jet.
%This coincidence increases the confidence in the reality of the
%apparent features.
%It is interesting to note that they are much more prominent in the
%spectral index than in the total intensity images. 

   There are several effects that will corrupt the spectral index
image.  The relatively poorer uv coverage at 3.6 cm wavelength will
result in poorer surface brightness sensitivity which will result in
incorrectly steep estimates of the spectral index.  The rapidly
declining surface brightness of the jet will cause any image
registration errors to result in systematic variations in the derived
spectral index.
What can be reliably determined from Fig. \ref{SpIndex} is that the
core region is optically thick and the jet spectral index steepens to
optically thin values.

\subsection{Polarization}

   Images of NGC~315 in linearly polarized light do not reveal any
polarized emission clearly due to the source, in spite of the low
noise level of 0.066 mJy/beam in the Stokes' Q and U images.
Off source apparent artifacts due to imperfect instrumental
polarization calibration do not exceed 0.22\% of the peak total
intensity in the image.

One possible cause of the lack of polarization would be a very high
rotation measure in the source in which variations across the bandpass
decorrelate the signal.
A rotation measure of 82,000 rad/m$^2$ will cause a 50\%
decorrelation across our full 32 MHz bandpass.
To test for the possibility of very high rotation measure in this
source, a polarized intensity image was generated for each of the 64
0.5 MHz channels in the data, and these polarization images were
averaged. 
This scalar averaging of the polarization results in a very slow increase in
signal-to-noise ratio with increasing number of channels (the final
sensitivity is only a few times that of a single 0.5 MHz channel) but
this process is much less sensitive to large rotation measures.
The half decorrelation rotation measure is 5,200,000 rad/m$^2$ for a
the scalar averaged polarization image.
No polarized emission was detected in the scalar averaged image; this
effectively eliminates the possibility of depolarization due to a
constant, but very large, Faraday rotation.
Table \ref{poln} gives upper limits to the source polarization  at a
number of locations along the jet using both the scalar and vector
averaged polarization images. 
The upper limit quoted is the polarized amplitude at the corresponding
image location. 
High resolution VLA images (to be published elsewhere) indicate 26\%
polarization at a distance of 700 mas from the core.

\begin{table} [t]
\begin{center}
\caption{Upper Limits to Polarization}
\medskip
\begin{tabular}{lrccll} 
\hline \hline 

Distance &  I  & $P_{scalar}$ & $P_{vector}$ & $\%_{scalar}$ & $\%_{vector}$ \\
  mas    & mJy &  mJy         &   mJy        &               &  \\
\hline
   0.0   & 381.5 & 0.99 & 0.18 & 0.26 & 0.05 \\
%% 0.0   & 381.5 & 0.99 & 0.18 & 0.25 & 0.05 \\
   5.5   &  33.1 & 1.0  & 0.10 & 3.02 & 0.30 \\
%% 5.5   &  33.1 & 1.0  & 0.10 & 3.3  & 0.3 \\
   10.25 &  9.4  & 1.1  & 0.29 & 11.7 & 3.09  \\
%% 10.25 &  9.4  & 1.1  & 0.29 & 11.2 & 0.3 \\
\hline
\label{poln}
\end{tabular}
\medskip \\
\end{center}  
\tablecomments{
Distance is from the peak of the core, I is the total intensity,
$P_{scalar}$ is the scalar averaged polarized intensity,
$P_{vector}$ is the vector averaged polarized intensity,
$\%_{scalar}$ is the scalar averaged percent polarization,
$\%_{vector}$ is the vector averaged percent polarization.
}
\end{table}

\section{Discussion}
\subsection{Polarization}

The low level of linear polarization from the inner region of this
radio source could be the result of a very disorganized magnetic field
in the jet in its inner few parsecs, Faraday depolarization inside the
jet, or Faraday depolarization in a screen in front of the jet.
An external Faraday screen must have large fluctuations on size scales
smaller than our resolution of about 1 pc (projected).
A constant, large Faraday rotation is effectively ruled out by the
lack of detected polarization averaged over many 0.5 MHz bands.
One possibility for an external depolarizing screen is the Narrow Line
Region (NLR). 
According to \cite{urry95} the NLR extends up to $\sim$ 32 pc,
corresponding to $\sim$ 70 mas for NGC315.
As described above, the jet is probably inclined to the line of sight
by approximately 35$^\circ$, so the line of sight to the inner portion
of the jet almost certainly intersects the NLR.
Much further from the core, at 700 mas, VLA observations show that the
jet is highly polarized (26\%); on this scale the jet is certainly
outside of the NLR.
Moreover, we have to take in account that the inner 5 - 6 mas are dominated
by the core region with an inverted spectrum (see Fig. 4) where the
intrinsic polarization is much lower than in the synchrotron transparent 
regions. 
High surface brightness sensitivity polarization observations at
higher frequencies are needed to determine the extent over which the
jet is unpolarized. 
Unfortunately, polarization VLBI observations of other FRI sources
are not yet available for a comparison.

\subsection{Nuclear Activity}
   The radio emission from the core of NGC~315 was relatively constant
during the 1970's when it was monitored by \cite{ekers83}.
Since the late 1980's, the core has been more active, with an event
apparently peaking in the early 1990's, although during these years the 
available flux density measurements are too sporadic to properly
define the light curve.  
%This outburst  may be related to the higher level of emission
%apparently propagating along the jet.
There is some fluctuation of the brightness of the unresolved ``core''
component, but most of the additional emission does not appear to be
associated with a distinct component. 
Although the measurements of the core variability  are too sparse to
allow a correlation with the evolution of the parsec scale structure,
they are  a clear indication of the activity in this radio galaxy.

\subsection{Proper motion}
   Comparing the images available for NGC~315 (Fig.
\ref{N315Cont}) at different epochs allows the detection of apparent motion.
The 1996 May image was not included due to the problems discussed above. 
The main difficulty in this analysis is that the jet is very
homogeneous and the evidence of substructure is not very strong.
%For the first 7 - 9 mas from the core we used the super-resolved images
%presented in Fig. \ref{N315HRCont}. 
%The good correspondence between the images at 3.6 and 6 cm shown in Fig.
%\ref{N315XCcomp} gives some confidence in the reality
%of the inner features in the super--resolved images shown in Fig.
%\ref{N315HRCont}. 
%The coincidence of spectral flattening with the apparent features in
%Fig. \ref{N315Cont} increases the reliability of these features,
%suggesting that they are physical rather than simple variations in the jet. 

\begin{figure}
\plotone{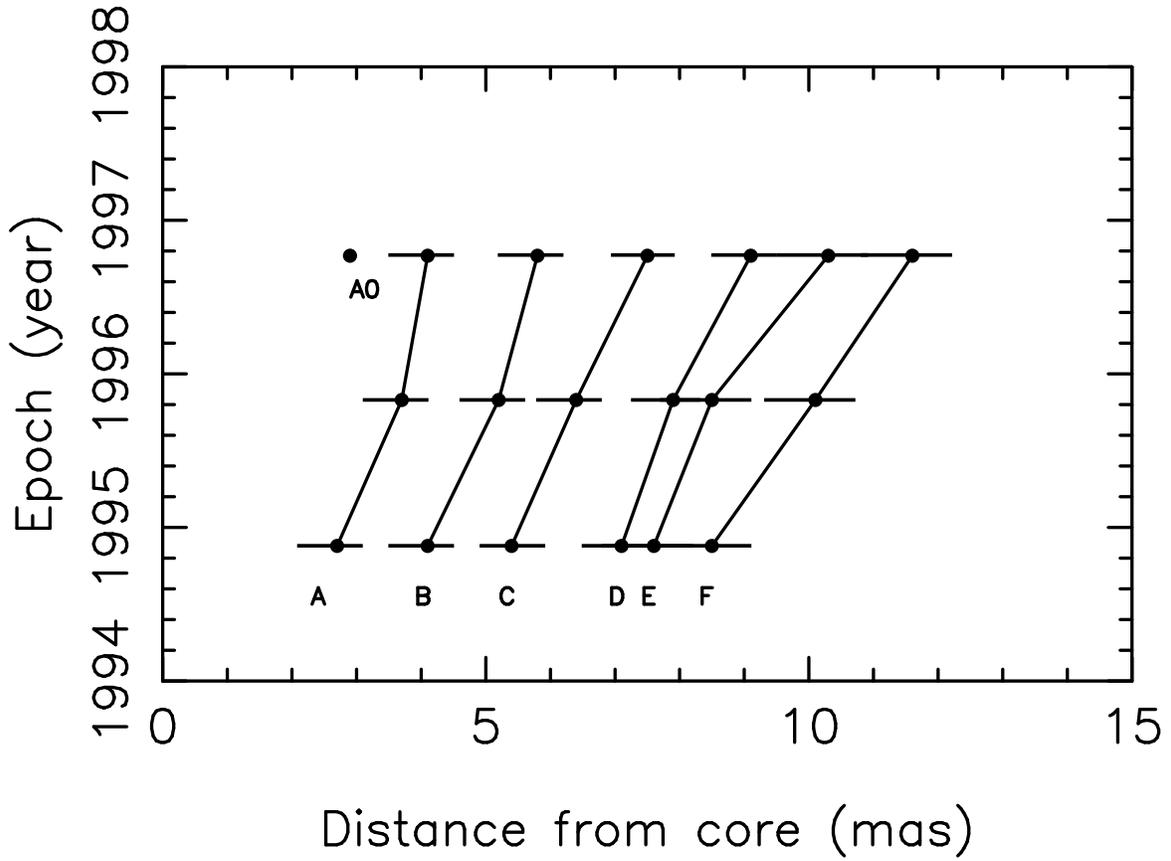}
\caption{
The locations of features along the jet in NGC315 at several epochs.
The vertical lines indicate the suggested associations.
Features A0, A, B, C and D are from the super--resolved images and E
and F from the images at normal resolution.
The horizontal lines give the approximate uncertainty of the location
of the feature.
\label{position}}
\end{figure}

 The distance from the core of the various features as a function of the
epoch is given in Table \ref{tabposition} and shown in Fig. \ref{position}. 
The inner four features in the first two epochs and the inner five in
the last epoch are from the super-resolved image (Fig.
\ref{N315HRCont}), while point E is from the normal resolution
images (Fig. \ref{N315Cont}).
The final feature, F, is the location of a sharp drop in the intensity
of the jet.
Feature F is not a {\it knot} in the usual sense but a well defined
position in the jet where there is a marked decline in emission,
perhaps associated with a new region of activity propagating along the
jet.
We are aware of the fact that the association of features at the different 
epochs shown by the lines in Fig. \ref{position} is not the only one
possible. 
However, it is the one which results in the best set of alignments.
In this interpretation of Fig. \ref{position}, feature A0 is newly
emerged from the core.

The measurements shown in Fig. \ref{position} are insufficiently
precise to show acceleration or deceleration of a given feature, but
the general steepening of the lines down the jet indicates an
acceleration of the jet.
The average velocity of each feature in the jet is used to
derive the $\beta _{app}$ given in Table \ref{tabposition}.
%The set of lines shown in Fig. \ref{position} suggests a clear proper
%motion with a possible acceleration along the jet. 

\begin{table} [t]
\begin{center}
\caption{Jet Feature Position and Apparent Velocity}
\medskip
\begin{tabular}{lccccccc} 
\hline \hline 

       &      A0  &     A   &    B   &    C  &    D  &     E  &   F \\
\hline
\\
1994 November  &   -      &    2.7  &   4.1  &   5.4 &   7.1 &     7.6&  8.5 \\
\\
1995 October  &    -     &    3.7  &   5.2  &   6.4 &   7.9 &     8.5& 10.1 \\
\\
1996 October  &     2.9  &    4.1  &   5.8  &   7.5 &   9.1 &    10.3& 11.6 \\
\\
\\
 $\beta_{app}$ &    - & 1.13$h^{-1}_{50}$ & 1.37$h^{-1}_{50}$ & 
                        1.70$h^{-1}_{50}$ & 1.62$h^{-1}_{50}$ & 
                        2.18$h^{-1}_{50}$ & 2.51$h^{-1}_{50}$ \\
\hline
\label{tabposition}
\end{tabular}
\medskip \\
\end{center}
\tablecomments{
Distance is from the peak of the 3.6 cm core in mas.
A0, A, B, C and D are from super-resolved images; E and F from normal
resolution images.
The apparent velocity ($\beta_{app}$) in c units is between
November 1994 and October 1996; h$_{50}$ is H$_0$/50.
}
\end{table}

\subsection{Jet/Counter Jet ratio}

\begin{figure}
\plotone{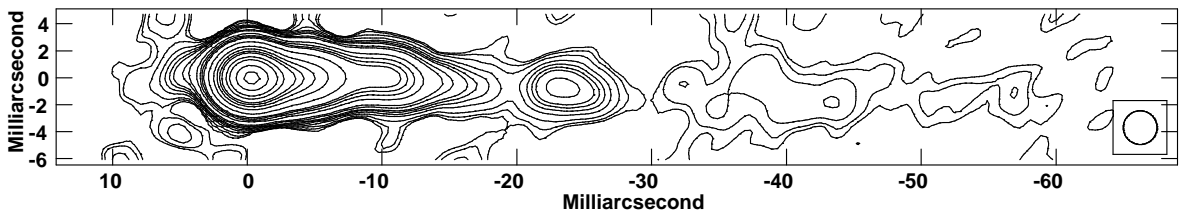}
\caption{
The naturally weighted image of NGC315 from the 1996 October data
showing the low brightness emission including the counter-jet.
The peak brightness is 436 mJy and contour levels are at 0.05 0.1,
0.2, 0.4, 0.6, 0.8, 1, 1.5, 2, 4, 6,  8, 10, 20, 40, 60, 80, 100, 150,
200, 300 and 400 mJy. The noise level is 0.06 mJy/beam.
The image has been rotated on the sky by -41.5$^\circ$;
the size of the restoring beam (2.5 mas) is shown in the lower right
corner. 
The position scale is relative to the 3.6 cm peak.
\label{N315Counterjet}}
\end{figure}

The high quality image of NGC~315 from 1996 October gives indication
of a faint counter-jet.
The faint and short structure  visible in the normal resolution 
(Fig. \ref{N315Cont}) and super-resolved (Fig. \ref{N315HRCont}) images
is even more evident when  natural weighting is used (Fig.
\ref{N315Counterjet}). 
%To test if this structure is real or not we tryed to take away it
%from the data cleaning only in boxes where the main jet was present
%and using these data to self-calibrate the uv-data.  
%However we were not able to take off this counter jet feature. 
%Moreover we note that it is evident in our image which was carefully
%calibrated reaching a very low noise level and very good zero level. 
%Moreover it is barely present in lower resolution images at the other
%epochs even if the higher noise level and the presence of residual in
%these lower quality images would not allow to believe to the reality
%of this structure. 
A test of the reality of this feature was to perform several
iterations of self--calibration disallowing CLEAN components in this
region.
The apparent counter-jet persisted throughout this procedure and
appears to be required by the data.
Including this region results in a map with very low noise level as well as a
decrease in the level of off--source artifacts.
Low resolution images at the other epochs are consistent with a
counter-jet, although the noise is too high to confirm its presence. 

   If the jets are intrinsically symmetric, the jet magnetic field has
a random orientation and the asymmetries are entirely due to Doppler
beaming, it is possible to determine the component of the jet velocity
in our direction using the relation: 
%Assuming that the jets are intrinsically symmetric, from the jet to
%counter--jet brightness ratio R we can constrain the value of $\beta$
%cos$\theta$, according to the formula:
$${\rm R} = [(1+\beta {\rm cos}\theta) / (1-\beta{\rm cos}\theta)]
^{2+\alpha}$$
where R is the jet/counter--jet brightness ratio,
$\beta$ is the ratio of the jet velocity to the speed of light, and
$\theta$ is the jet angle to the line of sight.
(see \cite{giovannini94} for a more detailed discussion). 
We note that NGC~315 shows increasingly symmetric, straight jets on
the kpc scale which supports the Doppler favoritism interpretation of
the pc scale asymmetries.

  A simple interpretation of the jet brightness asymmetries depends on
the jet being constant in time, otherwise the time delay between the
approaching and receding jets must included.
The ``features'' in the jet whose motions were derived in a previous
section are relatively minor fluctuations on the underlying jet so
they may be ignored for this analysis.

As discussed above, the recent flux density outburst in NGC~315
appears to have resulted in a general brightening in the inner jet and
a time invariant brightness ratio analysis must ignore the inner 4 mas
of the jet. 
The results of \cite{ekers83} suggest that there was an
extensive quiescent period prior to the recent outburst.

The image from 1996 October can be used to derive jet/counter--jet
ratios along the inner portion of the jet.
The inner 4 mas of the jet were avoided due the the very bright core
component and the recent outburst discussed above.
The spectral index was assumed to have a constant value of 0.5.
The measured brightness values are given in Table \ref{jcj} with the derived
values of $\beta$ cos$\theta$. 
If the angle to the line of sight $\theta$ is constant, the values
shown in Table \ref{jcj} imply a higher jet velocity, i.e. an
acceleration,  with increasing distance from the core.

   The derived values of $\beta$ cos$\theta$ given in Table \ref{jcj}
assumed a constant spectral index whereas Fig. \ref{SpIndex} suggests
a steepening of the spectrum along the jet.  
A steepening of the spectral index from 0.5 to 1.5 along the portion
of the jet summarized in Table \ref{jcj} could account for the
measured variations in jet/counter jet ratio from a constant velocity
jet.
As discussed above, the values of $\alpha$ shown in Figure \ref{SpIndex} are
subject to substantial error due to imaging difficulties.
The optically thick, inner portion of the jet has been excluded from
this analysis.
A jet spectral index of 1.5 this close to the core would require
implausibly rapid energy loss to the radiating electrons;  it seems
unlikely that the actual spectral steepening could be sufficiently
large to produce the increasing brightness ratios observed.

   Another possible cause of the jet/counter--jet assymetry could be
free--free absorption in an accretion disk around the nucleus as is
observed in NGC~1275 (\cite{walker98}).
%The absorbing disk in NGC~1275 appears to largely depolarize the jet
%emission. 
%Such a disk could be the source of the depolarization in NGC~315 as
%well as well as producing assymetries in the jet brightnesses.
However, the assymetries from free--free absorption should decrease
with distance from the nucleus as the optical depth through the
obscuration decreased. 
Since the observed jet/counter--jet ratio increases away from the core
it appears that free-free absorption is not a major contributor to the 
brightness assymetries.

\begin{table} [t]
\begin{center}
\caption{Jet Counter-jet brightness ratio}
\medskip
\begin{tabular}{lrrr} 
\hline \hline 
Core distance &  Jet Brightn. & CJ Brightn. & $\beta$ cos$\theta$ \\
  mas         & mJy/beam      & mJy/beam    &                     \\
\hline
4.0           &  75           &  1.80       & 0.63 \\
6.0           &  40           &  0.65       & 0.68 \\
7.5           &  19           &  0.15       & 0.76 \\
9.0           &  13           &  0.10       & 0.76 \\
10            &  13           & $<$ 0.06    & $>$0.79 \\
\hline
\label{jcj}
\end{tabular}
\end{center}
\end{table}

\subsection{Jet velocity}

\begin{figure}
\plotone{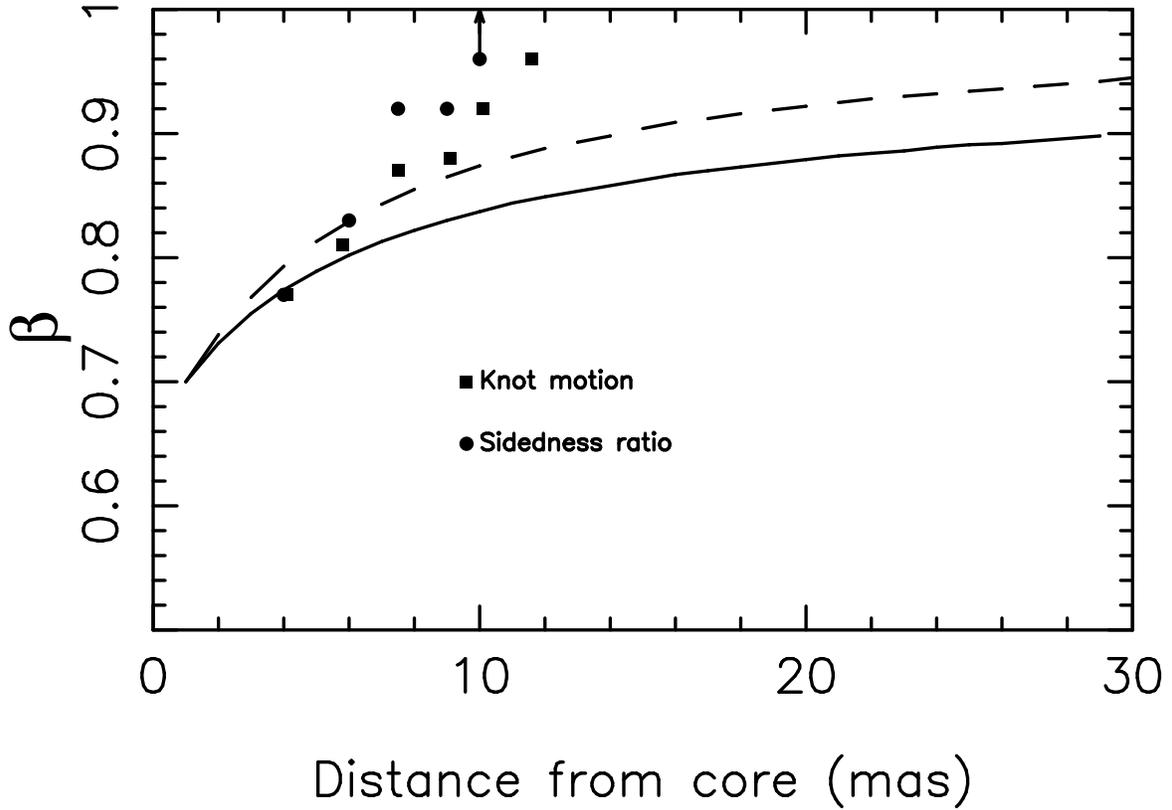}
\caption{
The derived jet velocities from apparent knot motions (squares) and
jet/counter--jet ratios (circles) at the October 1996 epoch.
The orientation to the line of sight is assumed to be 35$^\circ$,
H$_0$ = 50 and spectral index $\alpha$ = 0.5.
The lines show the jet velocity $\beta$ along the jet, obtained from
the same epoch, using the adiabatic model. 
The solid line was obtained assuming a magnetic field oriented
parallel to the jet axis, the dashed line uses the assumption of a
perpendicular field.  
\label{velocity}}
\end{figure}

The discussion above indicates that both the derived motion of
features in the jet and the analysis of the jet/counter--jet
ratio suggest that the jet is accelerating.
From the sidedness ratio (Table \ref{jcj}), we can derive the bulk jet
velocity assuming a reasonable orientation of NGC315 with respect to
the line of sight.
\cite{giovannini94}, have considered the jet orientation in NGC315
using a number of observational properties such as X-ray emission,
core dominance, and the very large linear size of this source as well
as the jet/counter--jet brightness ratio and concluded that $\theta$ is
in the range 30$^\circ$ to 41$^\circ$.
We adopt a value for $\theta$ of 35$^\circ$.
%Since the allowed range of $\theta$ is narrow for NGC315 
%(30$^\circ$ $<$  $\theta$ $<$ 41$^\circ$) as discussed in
%\cite{giovannini94}, we assume a value of $\theta$ = 35$^\circ$ for
%orientation of  the NGC315 jet. 
The derived values of the velocity are %given in Table \ref{jvel} and
plotted in Figure \ref{velocity}.

The apparent motion in units of the velocity of light for a
relativistically moving feature is given by the relationship:
$$\beta_{app} = \beta {\rm sin}\theta (1-\beta{\rm cos}\theta)^{-1}$$
Solving for $\beta$ gives:
$$\beta = \beta_{app}\times (\beta_{app} 
              {\rm cos}\theta+{\rm sin}\theta)^{-1}$$
Accurate determination of $\beta$ depends on knowing the jet's angle
to the line of sight as well as the value of H$_0$.
Using a value of 35$^\circ$ for the jet orientation we note that the
bulk and pattern jet velocity are very similar if we assume an Hubble
constant H$_0$ = 50 km sec$^{-1}$ Mpc$^{-1}$ (see Fig. \ref{velocity}).
This is consistent with the result of \cite{ghisellini93}, 
and the general agreement in the literature between the bulk and pattern
velocity for superluminal sources (see also \cite{giovannini98}).
In Table \ref{jvel}, we report the derived velocity from the visible
proper motion and the brightness ratio assuming H$_0$ = 50 km
sec$^{-1}$ Mpc$^{-1}$ and $\theta$ = 35$^\circ$ at different positions
in the jet.
The positions are de--projected linear distances in parsec from the
core assuming $\theta$ = 35$^\circ$.

\begin{table} [t]
\begin{center}
\caption{Jet Velocity}
\medskip
\begin{tabular}{cll} 
\hline \hline 

Core distance &  $\beta_{j/cj}$ & $\beta_{knots}$  \\
  pc         &                 &                   \\
\hline
3.3          &     0.77        &                   \\
3.4          &                 &    0.75           \\
4.8          &                 &    0.81           \\
4.9          &     0.83        &                   \\
6.2          &                 &    0.86           \\
6.2          &     0.92        &                   \\
7.4          &     0.92        &                   \\
7.5          &                 &    0.85           \\
8.2          &  $>$0.96        &    0.92           \\
9.5          &                 &    0.95           \\
%9.8          &  $>$ 0.96       &                   \\
\hline
\label{jvel}
\end{tabular}
\medskip \\
\end{center}
\tablecomments{
Distance is from the peak of the 3.6 cm core in pc de-projected for an angle of 
35$^\circ$ and using H$_0$ = 50.
$\beta$ is the jet velocity in c units derived using the brightness asymmetry
(Col. 2) or the visible proper motion (Col. 3).
}
\end{table}

These two, admittedly weak, derivations of the jet velocity give very
nearly the same values of jet velocity and the same acceleration along
the jet. 
Taken together, these two indicators of jet acceleration are stronger
evidence than either is individually.

If a value of H$_0$ = 100 is used, there is substantial disagreement
between the two putative indicators of jet velocity, while an intermediate
value of H$_0$ = 65 still gives comparable results. 
We note that, in any case, an increase in the jet velocity with the
distance from the core and a similar acceleration for the bulk and pattern
velocity is obtained.

Evidence of the presence of an accelerating jet is given in the
literature for a few other sources where jet substructures are found
to move at varying velocities and sometimes at varying angles with
respect to the line of sight.
In the radio galaxy 3C338, \cite{giovannini98} found a proper motion with 
an apparent velocity $\beta \sim$ 0.8 h$_{50}^{-1}$ at 4-5 mas from the core 
and a slower proper motion at 1.5 mas from the core. 
\cite{krichbaum98}, from two 1.3 cm observations of Cygnus A find evidence
for an apparent acceleration from $\beta_{app} \sim$ 0.2 to $\sim$ 1.2 
h$^{-1}_{50}$ in the region 1.5 - 3.2 h$^{-1}_{50}$ pc and finally
\cite{dhawan98} find  indication of acceleration along the parsec
scale jet in 3C84. 

In M87 (3C274), the parsec and kpc scale jet shows a complex
velocity field and no unambiguous conclusion can be drawn; however,
\cite{junor95} find that at very high resolution no proper
motion is visible and give a limit of 0.03c for the apparent jet
velocity in the inner 0.06 parsec. 
\cite{biretta95} found that velocities measured in the kpc scale
jet are much larger than  seen at the parsec scale. 
This evidence for an accelerating pattern speed may be in contrast
with the observed jet/counter--jet asymmetry in sources such as M87
(but not in 3C338).
Moveover, there are sources with  quasi--stationary features near the
core but with large jet/counter--jet ratios, suggesting a relativistic
bulk flow speed.
In NGC315, for the first time, there is evidence of an acceleration of
both the pattern and bulk flow velocities.
%Given the large jet/counter--jet asymmetry, the pattern speed in the
%inner jet of M87 must be less than the bulk flow speed.
%There are other sources with quasi stationary features near the core
%but with large jet/counterjet ratios, suggesting a relativistic bulk
%flow speed.
%In these sources an increase in the pattern speed does not necessarily
%imply an acceleration of the jet.
%The jet must either accelerate between these scales or the features
%seen with VLBI move much slower than the bulk flow speed. 
%At 1.5 kpc the M87 jet starts to decelerate and an apparent motion of 0.11c
%is found for knot 'C' (\cite{biretta95}). 
%Hough et al. 1996 and Zensus et al. (1995) studied the quasar 3C263
%and 3C345 respectively. 
%In each case a variable velocity was found;
%3C263 appeared to have a constant $\theta$ and increasing velocity
%while in 3C345 the apparent velocity increasing may be due to a change in the
%jet direction.

\subsection{An Adiabatic Expansion Model}

We have derived potential information about the jet dynamics by
applying the simple model in which the jet is adiabatically expanding.
Under this assumption, the jet velocity, brightness and radius
are related. 
The functional dependence between these parameters has been discussed
in the limit of non-relativistic bulk motion by \cite{fanti82,bicknell84,perley84}.
The case of relativistic bulk motion is considered by \cite{baum97},
who obtained the following relationships:  

Predominantly parallel magnetic field: $I_{\nu} \propto
(\Gamma_j v_j )^{-(2 \alpha +3)/3)} r_j^{-(10 \alpha
+9)/3)} D^{2+\alpha} $

Predominantly transverse magnetic field: $I_{\nu} \propto
(\Gamma_j v_j )^{-(5 \alpha +6)/3)} r_j^{-(7 \alpha
+6)/3)} D^{2+\alpha} $

\parindent 0pt
where I$_{\nu}$ is the jet surface brightness, $\alpha$ is the spectral index,
r$_j$, v$_j$ and $\Gamma_j$ are the jet radius, velocity and Lorentz 
factor, and D is the
Doppler factor: D = ($\Gamma_j(1-\beta cos\theta))^{-1}$.

\begin{figure}
\plotone{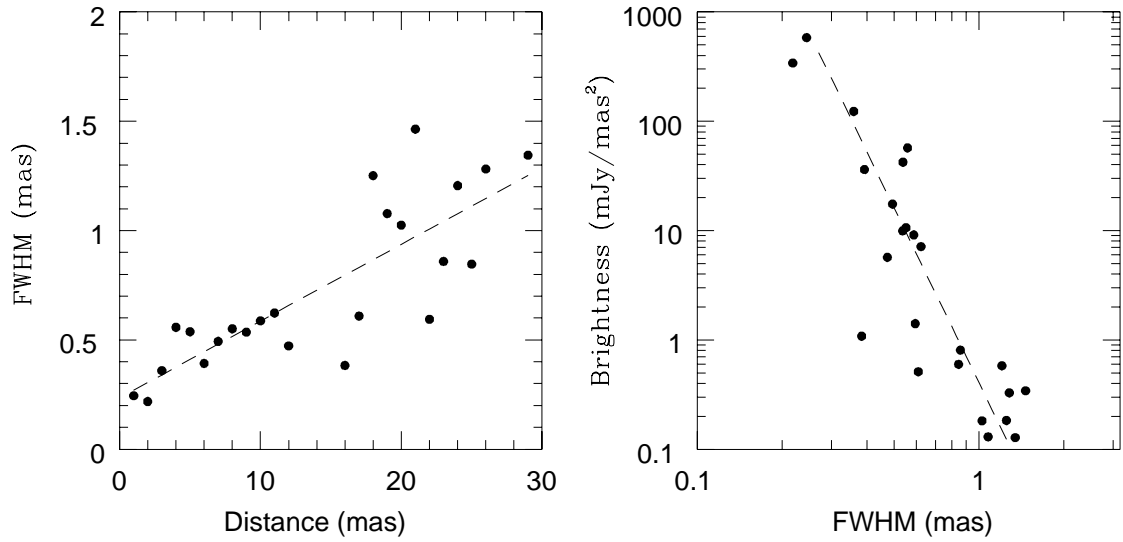}
\caption{
Left panel: plot of the deconvolved FWHM of the jet against the
distance from the core. The dashed line indicates the best fit to the
data, which has been used for fitting the adiabatic model.
Right panel: Plot of the deconvolved peak brightness  against 
the deconvolved FWHM of the jet. The dashed line shows the best fit
used for the adiabatic model. 
\label{Nr}}
\end{figure}

The jet transverse FWMH (full width at half maximum) and peak
brightness at increasing distance from the core were obtained by
fitting a Gaussian function to the transverse profiles of images at
resolutions of  2.5 mas (see Fig. 6) and 1.8 mas. 
The parameters were then deconvolved from the CLEAN beam, according to
the formula given by \cite{killeen86}, to get intrinsic quantities. 
The plot of these  parameters is given in Fig. 8.
The trend of the FWHM (left panel) shows a smooth increase with distance, with 
slope $\sim$ 0.035, corresponding to a  constant opening angle of 
$\sim$ 2$^{\circ}$. 
The best fit line does not contain the origin of the axes,
but gives at the core position (distance = 0) a FWHM of $\sim$0.2 mas.
In agreement with \cite{venturi93}, we believe that this is the
intrinsic angular size of the radio core. 
%We note that also \cite{krichbaum98} with higher resolution VLBA
%data, find that the minimum jet and counter-jet width in Cygnus A is
%$\sim$0.2 mas (note that Cygnus A is three times as distant as NGC~315).
The values of the  peak brightness decrease with the FWHM, according to
a FWHM$^{-5.31}$ law (right panel of Fig. 8). In an adiabatic jet
with constant velocity and $\alpha$=0.5, the trend implied by the above 
formulas  is I$_{\nu} \propto$ FWHM$^{-4.67}$ and  I$_{\nu} \propto$ 
FWHM$^{-3.17}$, for the parallel and transverse magnetic field,
respectively. 
The surface brightness observed in the jet decreases faster than can
be explained by adiabatic expansion; increasing Doppler dimming of an
accelerating, relativistic jet is one possible explanation.

Using the  relationships given above, we have modeled the jet
brightness and jet FWHM as a function of the distance from the core. 
To avoid the variations in the measured quantities due to local
fluctuations, or to the presence of blobs, we used the parameters
obtained from the best fits (see Fig. 8). 
We assumed a jet spectral index = 0.5,  a jet orientation to the line of sight
of 35$^{\circ}$, and a jet initial velocity (at 1 mas from the core) of 0.7c.
We considered the extreme possibilities of a purely parallel and
purely perpendicular magnetic field, since this information is not available
from the observations.
The derived values of $\beta$ are shown in Fig. \ref{velocity}. 
The jet velocity increases  along the jet up to the distance of 30
mas.
% Not really [WDC]:
% and is consistent  with the constraints obtained in the innermost 
%region by the apparent knot motion, and the jet/counter-jet ratios. 
The trend of the velocity is not crucially dependent on the
orientation of the magnetic field. 
Also, the use of a lower/higher initial velocity, does not change the overall
trend, but only scales it to lower/higher values. 

   The jet velocity derived from the adiabatic model is in reasonable
agreement with the other two estimates shown in Fig. \ref{velocity} up to a
distance of about 10 mas from the core, indicating an acceleration of
the jet.
After this point, the velocity derived from the adiabatic models are
lower than those from the other methods.
If this difference is meaningful, then it could indicate a
reacceleration of the relativistic particles in the jet in violation of
the assumption of an adiabatic jet.
The consistency of the adiabatic jet model with the derived proper
motion and brightness ratio velocities for the inner portion of
the jet further supports the accelerating jet interpretation of
the data.

%Not true, we can say anything about the jet being adiabatic:
%We can conclude that the adiabatic model provides a satisfactory fit
%to the observational data.

%\placefigure{Nr2}

\subsection{Where is the Observed Emission from?}

There is growing evidence that FRI jets on the kpc scale have highly
relativistic spines surrounded by lower velocity sheaths (\cite{laing96}).
The kpc scale structure of NGC~315 is consistent which such a model
and high quality images have been obtained for detailed modeling.

The situation on the pc scale is less clear.
A very highly relativistic spine would be invisible to us if the
orientation of the jet is indeed 35$^\circ$ from our line of sight and
only the slowest moving portions of a spine/sheath jet would be
visible. 
The high degree of depolarization of this source suggests a rich
environment for the jet providing material for entrainment and
acceleration by a highly relativistic spine.

The data presented here suggest an accelerating jet but doesn't
address the question of whether this is merely an acceleration of the
outer layers of a mostly invisible, highly relativistic jet or if the
jet has not yet accelerated to a highly relativistic state.
If we are seeing only the outer layers, the jet should appear to be
hollow and, observed with sufficient resolution, is easily
distinguished from a strongly center brightened filled jet.
Unfortunately, the surface brightness of the fully transversly
resolved portions of the jet is below the threshold of the images
presented here so this test cannot be performed.
Given the difficulties of accelerating a powerful jet many parsecs
from the central engine and lacking evidence to the contrary, it seems
most probable that the emission in the images in this paper come from
the outer layers of a highly relativistic jet.

\section{Conclusions}

   We have presented here a detailed study of the nuclear properties of NGC315.
We can conclude that:
 
a) The nuclear source shows occasional increases in the continuum radio
emission indicating very active periods. 
Unfortunately, the lack of a continuous monitoring does not allow a
comparison between the nuclear activity and variation in the parsec
scale morphology.
 
b) The 6 cm radio structure in the parsec scale jet of NGC315 is quite
smooth with some evidence of moving features and a faint counter--jet.
When interpreted in terms of a relativistic jet model, apparent motion
of the features in the jet, as well as the jet/counter--jet brightness
ratio, indicate an acceleration of the jet in its inner 3--10 parsecs.
A comparison of the jet's observed properties with the predictions of a
simple adiabatic expansion model 
%suggests that the jet regime is nearly adiabatic and 
further supports the interpretation of an increasing jet velocity.  
The present data is insufficient to determine if the observed
emission is from the entire jet or the slowest portions of an
otherwise invisible, highly relativistic jet.
 
c) The extremely low upper limits on the polarization of the jet indicate
either a very disorganised magnetic field in the inner parsecs of the
jet or Faraday depolarization either in the jet or in front of it.
A strong candidate for the depolarizing mechanism is Faraday
depolarization in the Narrow Line Region, future observations at
higher frequency with higher angular resolution will clarify this
point.

\acknowledgments

{\centerline { Acknowledgements}}
The authors would like to thank Alan Bridle, Roberto Fanti, Jose-Luis Gomez and
Daniele Dallacasa for several helpful discussions. 
We would also like to thank the staffs of the observatories who participated 
in these observations and the Socorro correlator staff.
L. F. and G. G. acknowledge partial financial support by the Italian
Ministry for University and Research (MURST) under grant Cofin98-02-32.

\clearpage

\clearpage

\begin{table}
\dummytable\label{tbl-3}
\end{table}

% This is the last table for this paper (as well as the first), so we
% should follow it with a \clearpage.  In order to force all the floating
% tables out of their buffers and onto vertical page lists, we must use
% \clearpage rather than \newpage. 

\clearpage

\clearpage

%\figcaption[fig9.ps]{
%Jet velocity $\beta$ along the jet, obtained from the October 1996
%epoch, using the adiabatic model. 
%The solid line was obtained assuming a magnetic field oriented
%perpendicularly to the jet axis, the dashed line uses the
%assumption of a parallel field. 
%The symbols are from Fig. 7.
%\label{Nr2}}

\end{document}